# Accurate super-resolution low-field brain MRI

Juan Eugenio Iglesias,[1,2,3,4] Riana Schleicher,[5] Sonia Laguna,[6] Benjamin Billot,[3] Pamela Schaefer,[2] Brenna McKaig, [7] Joshua N. Goldstein,[7] Kevin N. Sheth,[8†] Matthew S. Rosen,[1,2,9†] W. Taylor Kimberly[5†]

[†]These authors contributed equally to this work.

**Author affiliations:**

1 Athinoula A. Martinos Center for Biomedical Imaging, Massachusetts General Hospital and Harvard Medical School, Boston, MA 02114, USA

2 Department of Radiology, Massachusetts General Hospital and Harvard Medical School, Boston, MA 02114, USA

3 Centre for Medical Image Computing, Department of Medical Physics and Biomedical Engineering, University College London, UK

4 Computer Science and Artificial Intelligence Laboratory, Massachusetts Institute of Technology, Cambridge, MA 02139, USA

5 Department of Neurology and Center for Genomic Medicine, Massachusetts General Hospital and Harvard Medical School, Boston, MA 02114, USA

6 Swiss Federal Institute of Technology (ETH), Zurich, Switzerland.

7 Department of Emergency Medicine, Massachusetts General Hospital and Harvard Medical School, Boston, MA 02114, USA.

8 Department of Neurology, Yale New Haven Hospital, New Haven, CT 06511, USA

9 Department of Physics, Harvard University, Cambridge MA 02138 USA.

Correspondence to:
Juan Eugenio Iglesias
jiglesiasgonzalez@mgh.harvard.edu, and

W. Taylor Kimberly
Massachusetts General Hospital
Lunder 644
55 Fruit St, Boston, MA 02114, USA
wtkimberly@mgh.harvard.edu

**Running title**: Accurate super-resolution low-field brain MRI

**Abbreviations:** CNN = convolutional neural network; HF = high-field; HR = high-resolution; GMM = Gaussian mixture model; LF = low-field; LR = low-resolution; ML = machine learning; MPRAGE = magnetization-prepared rapid acquisition with gradient echo; QC = quality control; ROI = region of interest; SNR = signal-to-noise ration; SR = super-resolution

# Abstract

The recent introduction of portable, low-field MRI (LF-MRI) into the clinical setting has the potential to transform neuroimaging. However, LF-MRI is limited by lower resolution and signal-to-noise ratio, leading to incomplete characterization of brain regions. To address this challenge, recent advances in machine learning facilitate the synthesis of higher resolution images derived from one or multiple lower resolution scans. Here, we report the extension of a machine learning super-resolution (SR) algorithm to synthesize 1 mm isotropic MPRAGE-like scans from LF-MRI T1-weighted and T2-weighted sequences. Our initial results on a paired dataset of LF and high-field (HF, 1.5T-3T) clinical scans show that: *(i)* application of available automated segmentation tools directly to LF-MRI images falters; but *(ii)* segmentation tools succeed when applied to SR images with high correlation to gold standard measurements from HF-MRI (e.g., r = 0.85 for hippocampal volume, r = 0.84 for the thalamus, r = 0.92 for the whole cerebrum). This work demonstrates proof-of-principle post-processing image enhancement from lower resolution LF-MRI sequences. These results lay the foundation for future work to enhance the detection of normal and abnormal image findings at LF and ultimately improve the diagnostic performance of LF-MRI. Our tools are publicly available on FreeSurfer (surfer.nmr.mgh.harvard.edu/).

# Introduction

Magnetic resonance imaging has transformed the ability to evaluate the brain in both normal and pathological states. Conventional high-field MRI (HF-MRI) at 1.5-3T provides exquisite diagnostic resolution and contrast but these systems are costly and operate in access-controlled scanner suites that restrict imaging to patients who can be transported safely. Recently, we reported the first experience with a portable, low-field MRI (LF-MRI) 64mT system that can safely acquire neuroimages at the bedside of critically ill patients[1,2]. However, the innovative portability of an MRI scanner operating at 64mT comes with a penalty: decreased signal-to-noise ratio (SNR) and limited image resolution compared to HF-MRI. Therefore, a key goal for LF-MRI is to maximize the extraction of available information by transforming lower resolution LF-MRI data into higher resolution images while maintaining accuracy.

Quantitative morphometry measurements are central to many neuroimaging studies[3,4]. The ability to reliably extract quantitative measurements of the brain from LF-MRI would substantially extend its utility for research and clinical neuroimaging studies. There are a wide array of existing brain MRI segmentation tools used routinely, including FreeSurfer[5], FSL[6], and SPM[7]. Unfortunately, these tools have minimum prerequisites in terms of image resolution (e.g., isotropic 1 mm resolution), MR contrast (often T1-weighted scans), and SNR that are not met by LF-MRI scans, thus precluding application to these datasets.

Modern super-resolution (SR) methods use convolutional neural networks (CNN) to generate a high-resolution output from low-resolution input(s)[8]. In principle, this approach may enhance LF-MRI images sufficiently to be usable by standard neuroimaging tools[8,9], i.e., yield outputs with errors within acceptable margins. Supervised training of SR CNNs typically requires a large dataset of paired low-resolution and high-resolution images to learn a mapping from the former to the latter. However, such an approach would require very accurate alignment of the HF and LF scans for each subject, which is difficult due to nonlinear distortions. One common alternative is to downsample high-resolution scans to obtain paired images[8], but this approach is prone to failure in LF-MRI because downsampled HF-MRI scans do not closely resemble real LF-MRI images – a problem known as "domain shift"[10]. Recently, we developed a joint SR and contrast synthesis technique ("*SynthSR*") that uses synthetic data to overcome these limitations[11]. Notably, *SynthSR* can be trained to super-resolve MRI scans of any prescribed resolution and MR contrast – even in multi-modal scenarios with sequences of different voxel size and MR contrast. These features make the technique robust to the challenges of low resolution, domain shift, and imperfect co-registration.

In this study, we extend *SynthSR* so that it can utilize LF-MRI T1- and T2-weighted scans to generate an image with 1 mm isotropic resolution and MPRAGE contrast. We enhance the original *SynthSR* method by incorporating a segmentation-based regularizer, which further improves the quality of the output. Finally, using a test set of paired LF-MRI and HF-MRI scans, we show that our method enables the use of existing automated segmentation tools to generate brain morphometry measurements with high agreement between LF-MRI and HF-MRI scans.

# Materials and methods

## Training dataset

We used the HF-MRI dataset used to build the atlas for FreeSurfer's SAMSEG tool, which includes 20 MPRAGE scans with 1 mm isotropic resolution and corresponding segmentations of 39 regions of interest (ROIs): 36 brain ROIs, segmented manually, and 3 extracerebral ROIs, segmented automatically[12]. This dataset is also a subset of the training data used to build the atlas in FreeSurfer[5]. To improve the accuracy of tissue segmentation in the setting of pathological lesions (e.g., stroke, hemorrhage), we augmented the training dataset as follows. First, we subdivided the cerebral white matter into two components by fitting a Gaussian mixture model (GMM) to the underlying image intensities using the Expectation Maximization algorithm[13]. This created an additional class for abnormal image intensities in the training dataset, mostly due to white matter lesions. We then used this "lesion mask" to inpaint the underlying intensities using a publicly available, non-local means algorithm[14]. As described previously[14-16], inpainted images enhance the accuracy of subsequent automated segmentation.

## Super-resolution algorithm

The SR architecture builds on *SynthSR*[11], a supervised approach trained with: HF-MRI scans as target; and synthetic LF-MRI images derived from the segmentations of the HF-MRI scans as inputs. The crux of the architecture is the synthetic data generator (see Figure 1). At every iteration, this generator produces a mini-batch of data as follows. First, it randomly selects a real MPRAGE scan and its associated segmentation from the training data. Second, it spatially deforms them with a random nonlinear transform. Third, it samples synthetic images at high resolution using a bivariate GMM conditioned on the deformed segmentation. The parameters of the GMM are random, but also centered around values that generate T1-like and T2-like contrast (see Figure 1). Next, LF-MRI-like synthetic scans are obtained by downsampling the high-resolution synthetic scans to the resolution of our standard LF-MRI sequences (1.6×1.6×5.0mm) and adding Rician noise. Finally, the LF-like scans are upscaled back to high resolution (which simplifies further processing by letting all images live in the same voxel grid) and a random, smooth, non-negative bias field is applied to each channel (T1 and T2). The bias fields were obtained by sampling a low-dimensional (4×4×4) volume of independent Gaussian variables centered at zero; upscaling it to the dimensions of the HR scans; and taking voxel-wise exponentials, which serve two purposes: ensuring nonnegativity, and assigning the same probability to multiplying and dividing by the same factor.

We then applied this synthetic data generator to train a CNN to recover a HF-MPRAGE scan from the synthetic LF-MRI-like T1- and T2-weighted scans (see Figure 2). Most modern SR architectures assume that the LR input is a blurry version of the HR target, which is not the case in our application, due to the higher noise levels in the LF scans, as well as the differences in MR contrast between the LF T1 intensities and their HF MPRAGE counterparts. Instead, we use a 3D U-net[17,18] to solve the contrast synthesis and SR simultaneously[11]. The inputs of the U-net comprise the simulated LF T1 and T2 scans, and the output is the synthetic 1 mm isotropic MPRAGE. In *SynthSR*, this U-net is trained by comparing the ground truth and estimated MPRAGE volumes and minimizing the sum of absolute errors (henceforth "intensity loss", $\mathcal{L}_I$; see top half of Figure 2).

Here, we further enhance the synthesis by complementing the intensity loss with a second loss function based on a pre-trained segmentation U-net as regularizer (see bottom half of Figure 2). This second U-net has been trained to segment the HF-MRI scans using the original intensities and labels directly, and its weights are frozen while training the SR U-net. Its role is the following: at every mini-batch, the output of the SR U-net is not only compared to the ground truth, but also fed to the segmentation U-net. The estimated labels are compared with the ground truth segmentation using a Dice loss[19] (henceforth "segmentation loss", $\mathcal{L}_S$). This loss ensures that the synthesis is accurate in regions where subtle intensity differences that do not heavily increase the intensity loss may lead to large segmentation mistakes, e.g., the lateral regions of the thalamus, or the globus pallidum. The full expression for the loss $\mathcal{L}$ that is used to train the network is finally:

$$\begin{aligned}\mathcal{L} &= \mathcal{L}_I + \lambda\,\mathcal{L}_S \\ &= \frac{1}{|\Omega|}\sum_{x\in\Omega}|R_I(x) - P(x;\theta)| - \lambda\,\frac{1}{L}\sum_{l=1}^{L} Dice\big[(R_S == l), Seg\big(P(x;\theta)\big) == l\big], \qquad (1)\end{aligned}$$

where $\lambda = 0.25$ is a relative weight; $\Omega$ is the image domain; $x$ represents spatial location, $R_I$ and $R_S$ are the reference (ground truth) HF MPRAGE intensities and segmentation, respectively; $P(x;\theta)$ is the image intensity predicted by the SR U-net at location $x$ when its weights are equal to $\theta$; $L$ is the number of unique labels, and *Seg()* represents the action of the segmentation network with frozen weights. Implementation details of the CNN can be found in the supplement.

## LF-MRI / HF-MRI test cohort

The test cohort was derived from an ongoing, prospective observational study performed from August 2020 to December 2021 at Massachusetts General Hospital in the Neurointensive care unit and Emergency department. Patients presenting with neurological symptoms and who were scheduled for standard of care neuroimaging (MRI or CT) were approached for participation. Consent was obtained for all patients under the approval of the local Institutional Review Board. Exclusion criteria included an inability to lay flat, a body habitus preventing LF-MRIg, or the presence of MRI contraindications (cardiac pacemaker, insulin pump, deep brain stimulator, or cochlear implants). Those consented subjects with paired HF-MRI scans who did not have strokes were included in this analysis (n=11). A 64mT portable LF-MRI (Hyperfine Swoop) was used to acquire all LF scans. Details on the scanning protocols, pulse sequences and parameters are included in the supplement.

Gold standard segmentations – and associated ROI volumes – were obtained by processing the HF-MRI scans with our learning-based approach *SynthSeg*[20,21]. This tool is robust against variations in orientation, resolution, and MRI contrast, and can thus handle the heterogeneity in the acquisition of the HF-MRI clinical scans.

## Synthesis, super-resolution, and segmentation of LF-MRI

The LF-MRI data were processed by co-registering the T1- and T2-weighted LF-MRI scans using *NiftyReg*[22]. The co-registered LF-MRI scans where then processed with the trained *SynthSR* CNN to obtain the synthetic 1 mm MPRAGE output. The synthetic MPRAGE scans were subsequently segmented with *SynthSeg*[20,21], which is robust against mistakes in SR. We note that, in most cases, the field of view of the LF-MRI scans did not fully cover the cerebellum, so we left this ROI out of subsequent analyses.

## Data availability

The code for training the CNNs is publicly available on GitHub (https://github.com/BBillot/SynthSR). The trained model and inference code is distributed with FreeSurfer (https://surfer.nmr.mgh.harvard.edu/), such that anybody can easily download FreeSurfer and process their LF scans with a simple command:

```
mri_synthsr_hyperfine --t1 [t1_input] --t2 [t2_input] --o [output_file].
```

The LF-MRI imaging data are available from the corresponding authors upon reasonable request and in accordance with institutional data sharing agreements.

# Results

We first generated synthetic 1mm isotropic MPRAGE scans from the LF-MRI T1- and T2-weighted, low-resolution inputs. Examples from three representative cases are shown in Figure 3, which includes LF-MRI (64mT), synthetic MPRAGE, and the corresponding real HF-MRI MPRAGE – displayed in coronal, sagittal, and axial view. Next, to evaluate the performance of *SynthSR* model, we compared the ability of several segmentation tools to generate ROI volumes relative to the gold standard volumes obtained from the HF-MRI clinical scans. We assessed the ROI volumes generated by four publicly available segmentation tools (*FreeSurfer*[5], *SynthSeg*[20,21], *FSL-FIRST*[6], and *SAMSEG*[12]) directly applied to the LF-MRI images, and our proposed extension of *SynthSR*. We note that, like *SynthSR*, *SAMSEG* can jointly exploit the information in the T1 and T2 LF-MRI scans; *FreeSurfer* and *FSL-FIRST* can only use the T1 scans; and *SynthSeg* can only use one scan at the time – either then T1 or the T2. Examples of these segmentations are also shown in Figure 3.

Table 1 shows the correlation between the gold standard volumes (derived from the HF-MRI scans) and the volumes estimated by the different competing methods from the LF-MRI data; more detailed scatter and Bland-Altman plots for each brain ROI can be found in Supplementary Figures 1-3. *FreeSurfer* and *SynthSeg* were unable to produce usable segmentations of the LF-MRI scans and did not yield significantly correlated ROI volumes when compared to the gold standard. FSL-FIRST and SAMSEG were able to generate usable segmentations in some ROIs, with volumes that were moderately correlated with the gold standard (Table 1 and Supplementary Figures 1-2). By comparison, *SynthSR* was able to generate volumes that were much more highly correlated with the ground truth: the correlation was very strong ($r>0.8$) for hippocampus, thalamus, ventricles, white matter, and cortex; and moderately strong ($r>0.6$) for amygdala, caudate, putamen, and pallidum (Table 1 and Supplementary Figure 3).

The Bland-Altman plots also show that *SynthSR* exhibits reduced bias: *SynthSR* has minor biases in the pallidum, ventricles, and thalamus, while FSL-FAST displays biases for nearly every structure, and SAMSEG shows biases for the cortex, hippocampus, caudate, putamen, ventricles, and thalamus.

Comparing correlation coefficients across methods with a Steiger's test for dependent correlations[23] shows that our method produces significantly higher correlations for the amygdala and pallidum (compared with FSL-FIRST); caudate and putamen (compared with SAMSEG); and hippocampus and thalamus (compared with both). *SynthSR* did not produce correlations significantly lower than SAMSEG of FSL-FIRST for any of the ROIs.

# Discussion

While recent advances in brain MR image analysis and assembly of a corpus of well-annotated training images facilitate segmentation, quantitative analysis, and pathology detection[24,25], application to LF-MRI is currently hindered by the intrinsically lower SNR and resolution of the images. For example, a well-established subcortical segmentation technique like FSL-FIRST struggles to find the boundaries of the ROIs due to the poor resolution and contrast of the input data, rendering the output unusable (see examples in Figure 3). A similar effect is observed with another well-established technique that uses Bayesian segmentation: while SAMSEG's detailed probabilistic atlas and ability to exploit joint T1-T2 information enable the method to produce strong correlations for some structures (amygdala, pallidum, and larger structures like the cortex, white matter, and ventricles), the correlations for the rest of ROIs (hippocampus, caudate, putamen, and thalamus) are weak and not significant. Qualitatively speaking, the SAMSEG segmentation of the cortex fails to follow the gyri and sulci, and many structures are incorrectly segmented due a lack of contrast and resolution that the atlas cannot cope with (e.g., the putamen is incorrectly segmented across all examples in Figure 3).

Our proposed approach, on the other hand, holds promise to increase the image quality of LF-MRI scans to the point that they are usable not only by automated segmentation methods, but also registration and classification algorithms. Despite the limited resolution and contrast of the LF-MRI scans, *SynthSR* is able to produce a synthetic MPRAGE scan with sufficient detail and contrast for an automated segmentation algorithm to produce reliable volume estimates. Even though the synthetic scans are not as crisp as the ground truth MPRAGEs (Figure 3), and despite the fact that some ROI volumes are under or overestimated slightly, the segmentations of the synthetic and authentic scans are very highly correlated – even for convoluted structures such as the cerebral cortex. The quantitative results further demonstrate that the volume correlations are significant for *all* ROIs, with very strong correlations for the large ROIs (cortex, white matter, ventricles, whole brain), hippocampus, and thalamus; and strong correlations for all other structures.

Our study has several limitations. First, the sample size is small, due to the limited number of subjects with paired LF/HF-MRI data, however the use of a real-world clinical population increases the generalizability of our findings. Second, the validation is limited to a single downstream task (segmentation), which is a quantitative proxy for the ultimate downstream task: detecting clinically relevant findings. And third, our method does not currently estimate the uncertainty in the output, which may differ in regions with low image contrast or in spatially convoluted neuroanatomy. Incorporating uncertainty estimates is an area of future research that will further increase the reliability of our method .

Taken together, our approach provides a critical foundation for exploiting the benefits of deep-learning based reconstruction. As LF-MRI becomes increasingly available, methods like *SynthSR* have the potential to augment the impact of LF-MRI in neuroimaging.

# Implementation details

We used a U-net architecture (both for SR and regression) that we have successfully used in our previous work with synthetic MRI scans[20]. In short, it is a 3D U-net with five levels. Each level has two layers, each comprising convolutions with 3×3×3 kernels and an ELU nonlinearity[26]. The first layer has 24 kernels. The number of kernels is doubled after each max-pooling and halved after each upsampling. The last layer uses a linear activation to produce the estimated image intensities. The U-nets and the synthetic data generator are concatenated into a single model that is fully implemented on the graphics processing unit (GPU) using Keras (http://keras.io) with a Tensorflow backend.[27]

The hyperparameters that govern the GMM were estimated from a separate sample of five Hyperfine scans as follows. First, we segmented them automatically with SAMSEG[12]. Albeit very noisy, this segmentation enables rough estimation of means and variances, which we performed with robust statistics: we estimated means with medians, and standard deviations with median absolute deviations (multiplied by 1.4826, as per *Leys et al.*[28]). Crucially, we multiplied the standard deviations of all parameters by a factor of *5.0*, in order to provide the CNN with a considerably wider range of intensity distributions than we expect to see at test time. This factor makes the CNN more robust against variations in acquisitions (i.e., the domain gap), while also mitigating the impact of segmentation errors made by SAMSEG.

During training, images were min-max normalized and randomly cropped to 160×160×160 volumes; such cropping enables training on smaller GPUs. We set the learning rate to $10^{-4}$, and trained the CNNs for 200,000 iterations, which was sufficient for convergence (the training loss flattened at about 100,000 – 150,000 iterations). Training took approximately a week on a RTX6000 GPU; inference, on the other hand, takes about three seconds on the same GPU, and 10 seconds on a modern central processing unit (CPU).

## MRI scanning protocols

The LF-MRI imaging protocol included deep learning reconstructed T1-weighted, T2-weighted, and FLAIR scans; and a conventionally reconstructed diffusion-weighted sequence. All images were acquired with axial orientation. The T1 parameters were: acquisition time = 4:30 min, TR/TE = 1500/5.5 ms, TI = 300 ms, resolution = 1.6x1.6x5 $mm^3$, 36 slices, 5 mm slice thickness. The T2 parameters were: acquisition time = 4:32 min, TR/TE = 2000 /218 ms, resolution = 1.5x1.5x5 $mm^3$, 36 slices, 5 mm slice thickness. The FLAIR and diffusion-weighted images were not used in this study.

For the HF-MRI, clinical scans were obtained as follows:

- Three subjects: 3D sagittal MEMPRAGE scans, TE=1.69, TR=2530, TI=1300ms, echo number=1, Flip angle=7, resolution=1mm isotropic, acquired on a 3T Siemens Skyra.
- Two subjects: 2D sagittal SPGR T1-weighted scans, TE=2.46ms, TR=240ms, echo number=1, Flip angle=80, slice thickness=4mm, in-plane resolution=0.94mm, acquired on a 3T Siemens Skyra.
- Six subjects: 2D sagittal IR T1-weighted scans, TE=26.2, TR=2667.7ms, TI=808.5, Flip angle=160, slice thickness=5mm, in-plane resolution=0.47mm, acquired on a 1.5 GE Signa Artist.

# Funding

This study was supported by funding from the American Heart Association (Collaborative Science Award 17CSA3355004, co-PI KS, MSR, and WTK). The Hyperfine device was provided to Massachusetts General Hospital as part of a research agreement. KNS is supported by the NIH (U24NS107136, U24NS107215, R01NR018335, U01NS106513) and the American Heart Association (18TPA34170180, 17CSA33550004) and a Hyperfine Research, Inc. research grant. WTK is supported by the NIH (R01 NS099209 and R21 NS120002), and the American Heart Association (17CSA33550004 and 20SRG35540018). JEI is supported by the NIH (1R01AG070988), the BRAIN Initiative (1RF1MH123195), the European Research Council (Starting Grant 677697, project "BUNGE-TOOLS") and Alzheimer's Research UK (Interdisciplinary Grant ARUK-IRG2019A-003).

# Competing interests

W.T.K. has received research grant support from Hyperfine, Inc. W.T.K. also reports grants and personal fees from Biogen, and from NControl Therapeutics outside the submitted work; in addition, Dr. Kimberly has a patent to PCT/US2018/018537 pending and licensed. K.N.S. has received research grant support from Hyperfine, Inc. M.S.R. is a founder and equity holder of Hyperfine, Inc. The other authors declare no competing interests relevant to this project. The funding sources had no role in the design and conduct of the study; collection, management, analysis, and interpretation of the data; preparation, review, or approval of the manuscript; and decision to submit the manuscript for publication.

# Tables

**Table 1 Pearson correlation of ROI volumes obtained from HF-MRI (ground truth) and LF-MRI (with FSL-FIRST, SAMSEG, and our method)**

| Method / ROI | FSL-FIRST | | SAMSEG | | Proposed | |
|---|---|---|---|---|---|---|
| | **r** | **p-value[a]** | **r** | **p-value[a]** | **r** | **p-value[a]** |
| Cerebral-cortex | N/A | N/A | 0.92 | $< 10^{-4}$ | 0.93 | $< 10^{-4}$ |
| White matter | N/A | N/A | 0.90 | $< 10^{-3}$ | 0.93 | $< 10^{-4}$ |
| Hippocampus | 0.40 ** | 0.23 | 0.2 ** | 0.54 | 0.85 | $< 10^{-3}$ |
| Amygdala | 0.25 * | 0.46 | 0.74 | 0.009 | 0.66 | 0.03 |
| Caudate | 0.62 | 0.04 | 0.05 * | 0.88 | 0.71 | 0.01 |
| Putamen | 0.46 | 0.16 | 0.24 * | 0.47 | 0.62 | 0.04 |
| Pallidum | 0.25 ** | 0.47 | 0.81 | 0.002 | 0.71 | 0.01 |
| Ventricles | N/A | N/A | 0.95 | $< 10^{-5}$ | 0.97 | $< 10^{-6}$ |
| Thalamus | 0.51 ** | 0.11 | 0.4 * | 0.21 | 0.84 | 0.001 |
| Whole-cerebrum | N/A | N/A | 0.91 | $< 10^{-4}$ | 0.92 | $< 10^{-4}$ |

[a] P-values are for two-tailed t-tests assessing whether the correlation coefficient is significantly different from zero.

*$P < 0.05$, **$P < 0.01$, for Steiger tests for dependent correlations, assessing whether the correlation coefficients for FSL-FIRST or SAMSEG were significantly lower than those for our method. We note that our method did not produce correlations significantly lower than SAMSEG of FSL-FIRST for any of the ROIs.

# Figures

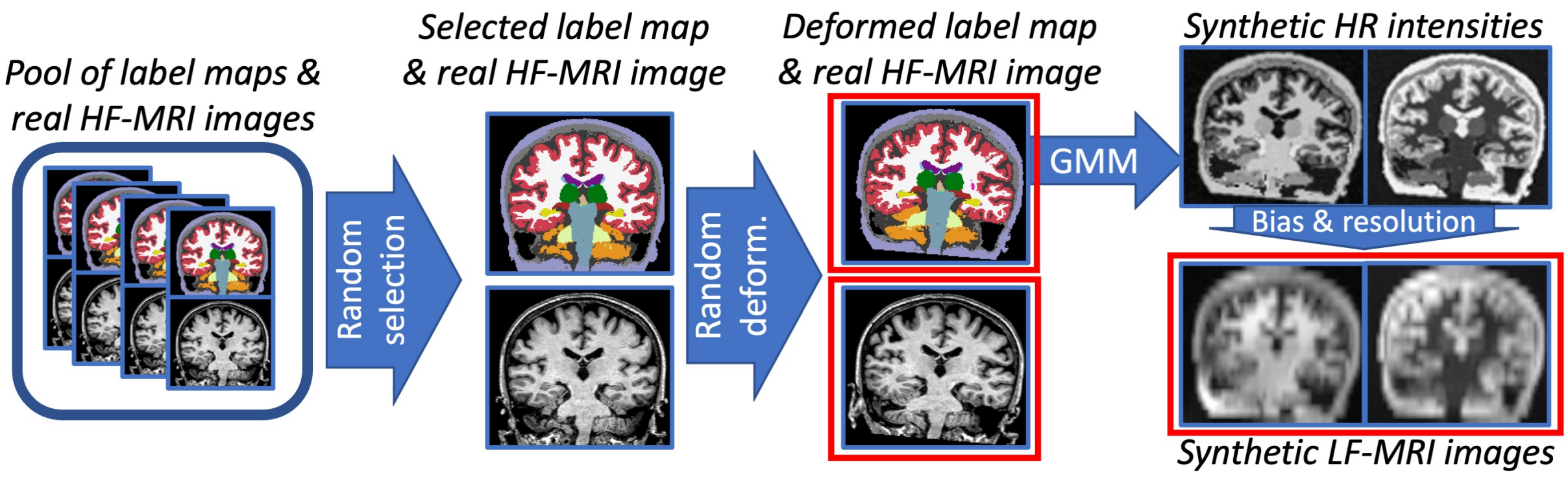

**Figure 1 Synthetic data generator.** Our method uses a very flexible generator to sample synthetic images on the fly during training. Using a pool of 3D segmentations and real HF images (the training dataset), it applies a combination of spatial augmentation, GMM sampling, resolution modeling, and bias field simulation to generate mini-batches with the following image volumes (in red boxes): deformed HF (HR) intensities, deformed HR segmentations, and simulated LF images with T1- and T2-like contrast.

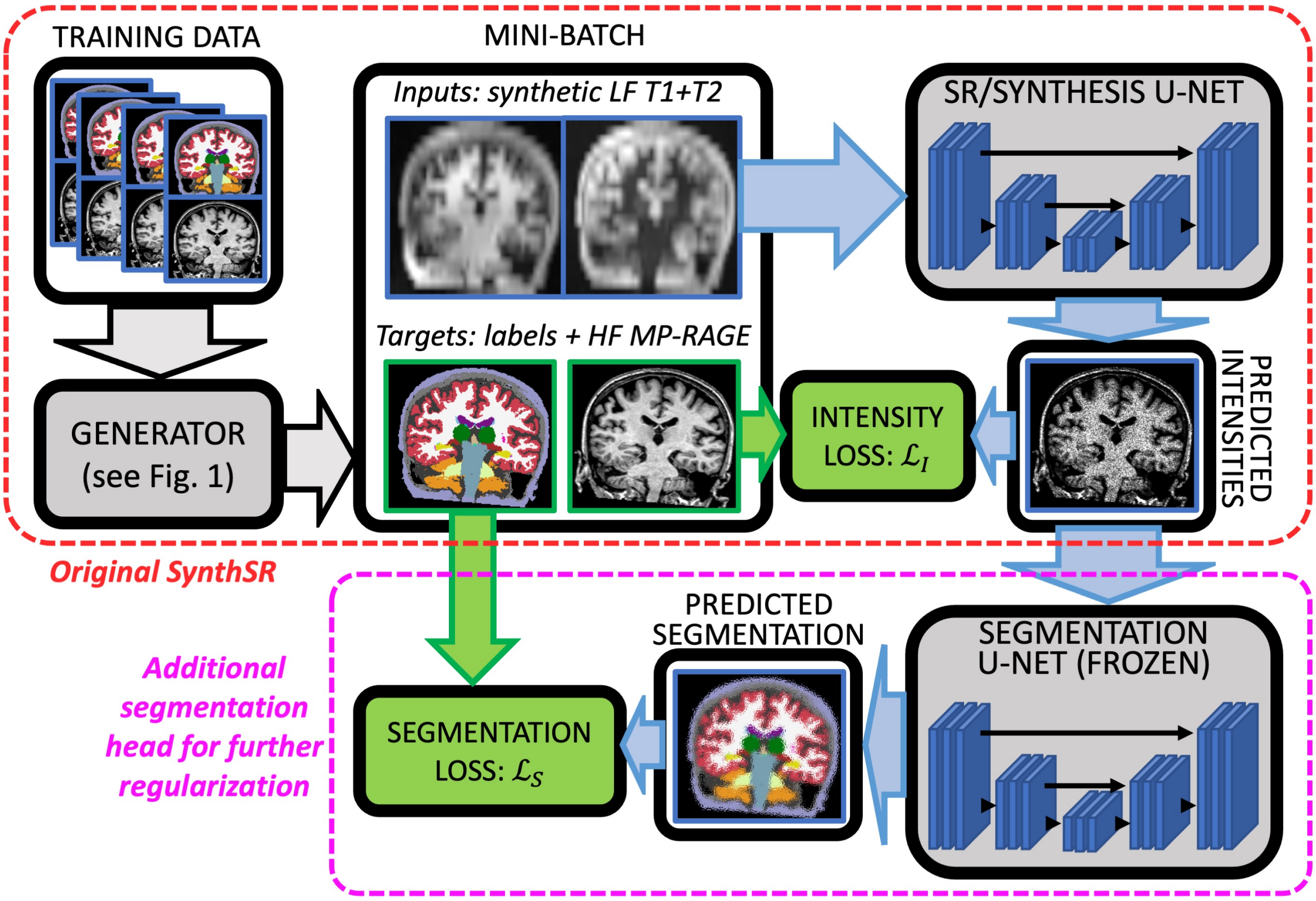

**Figure 2 CNN architecture.** The generator in Fig. 1 uses the training dataset to produce mini-batches on the fly, which consist of synthetic LF images and corresponding HF images and HR segmentations. The SR U-net predicts HR intensities that are compared to the ground truth HF image to update the U-net weights. In addition, the predicted intensities are further fed to a segmentation U-net pre-trained to segment 1 mm MPRAGE scans (with frozen weights), and the output is compared with the ground truth segmentation to inform training. We note that the upper half of the figure (inside the red dotted line) corresponds to the architecture in the original *SynthSR* paper; in this work, we have added the segmentation head (inside the purple dotted line) for increasing the accuracy of the synthesis.

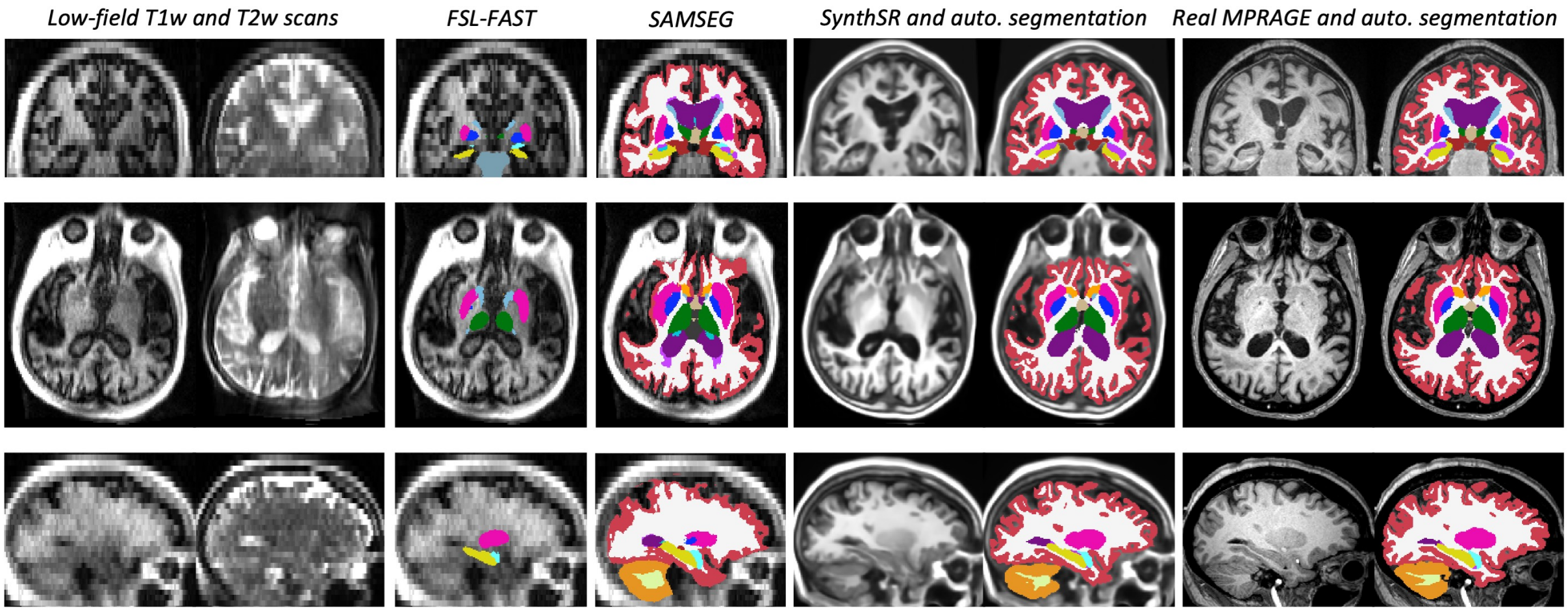

**Figure 3 Qualitative comparison of SynthSR with automated segmentation tools.** Each of the three rows corresponds to a different case and shows corresponding slices of: the LF T1-weighted and T2-weighted scans; the segmentations produced by FSL-FIRST and SAMSEG; the output of *SynthSR* and its automated segmentation with *SynthSeg* (third column); and the registered real HF MPRAGE and its automated segmentation with *SynthSeg* (fourth column). The slices are coronal for the first example, axial (which is the native orientation of the LF-MRI scans) for the second, and sagittal for the third.

# Supplementary figures

*FSL-FAST*

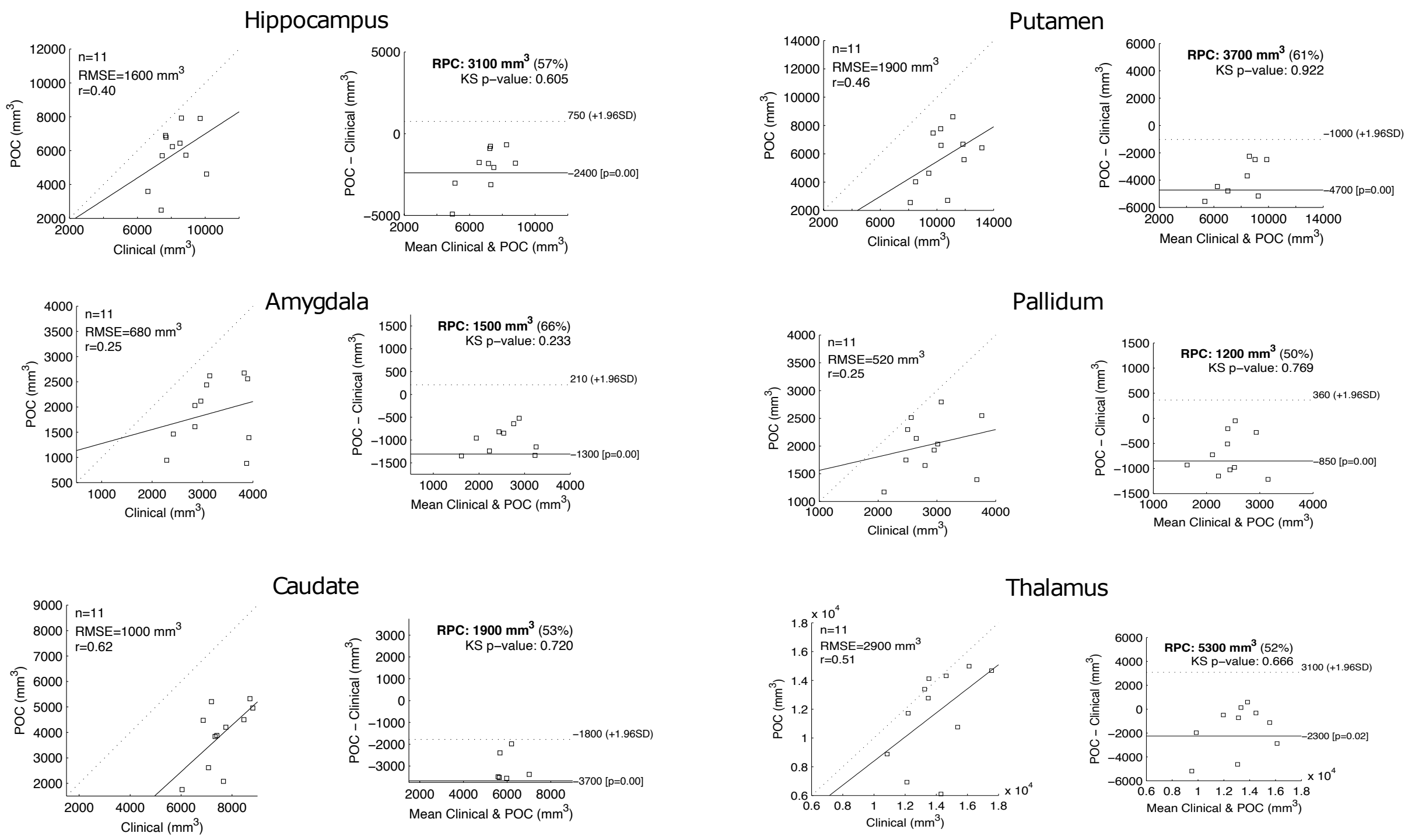

**Supplementary Figure 1 Scatter and Bland-Altman plots for FSL-FIRST.** Scatter (left) and Bland-Altman plots (right) comparing the ROI volumes derived from the HF-MRI clinical scans (ground truth) and from the LF-MRI scans using FSL-FIRST. In the Bland-Altmann plots, RPC stands for reproducibility coefficient, and the KS p-value is for a Kolmogorov-Smirnov test of normality of the differences.

*SAMSEG*

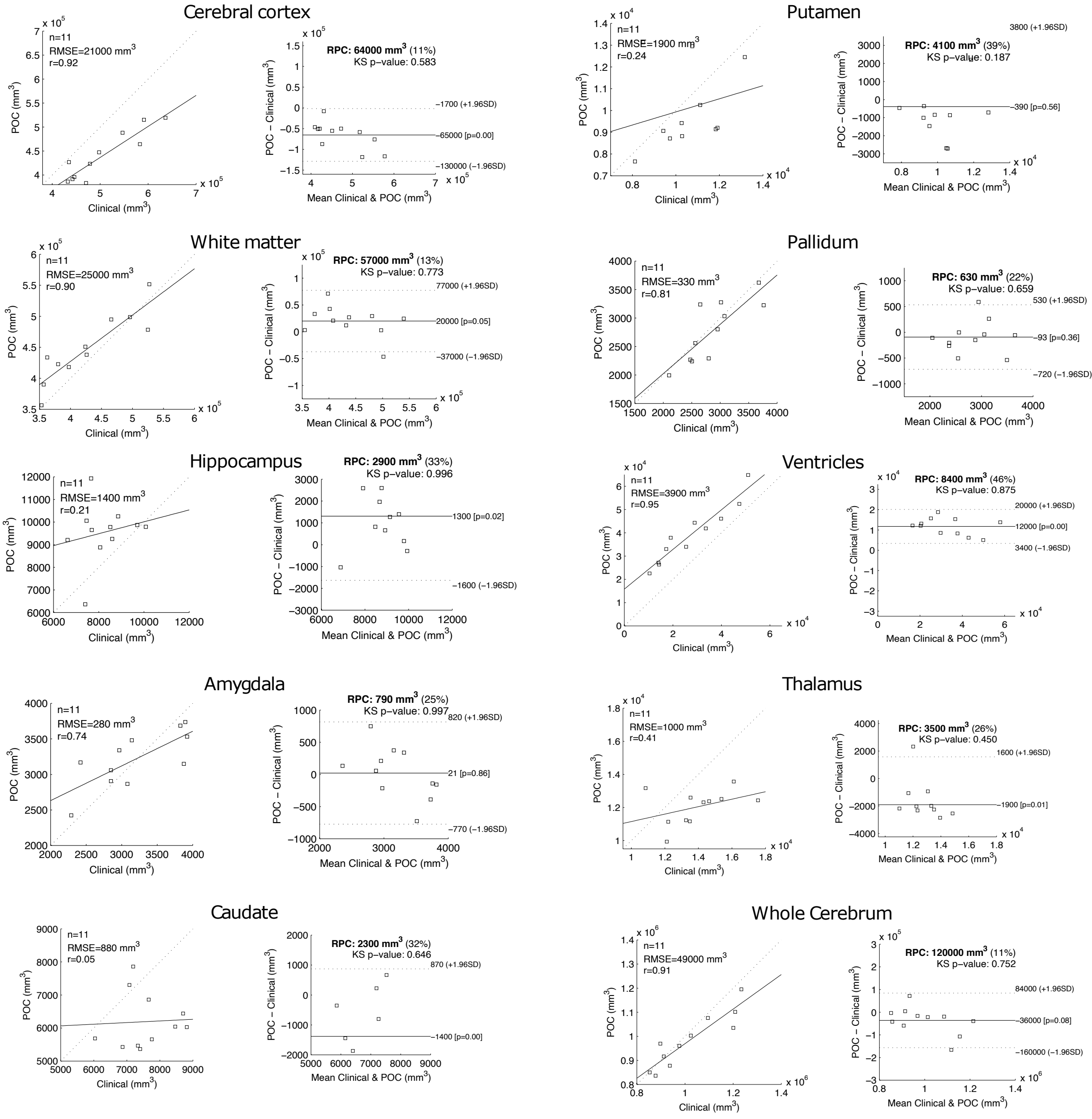

**Supplementary Figure 2 Scatter and Bland-Altman plots for SAMSEG.** Scatter (left) and Bland-Altman plots (right) comparing the ROI volumes derived from the HF-MRI clinical scans (ground truth) and from the LF-MRI scans using SAMSEG. Please see caption of Supplementary Figure 1 for further details.

*SynthSR*

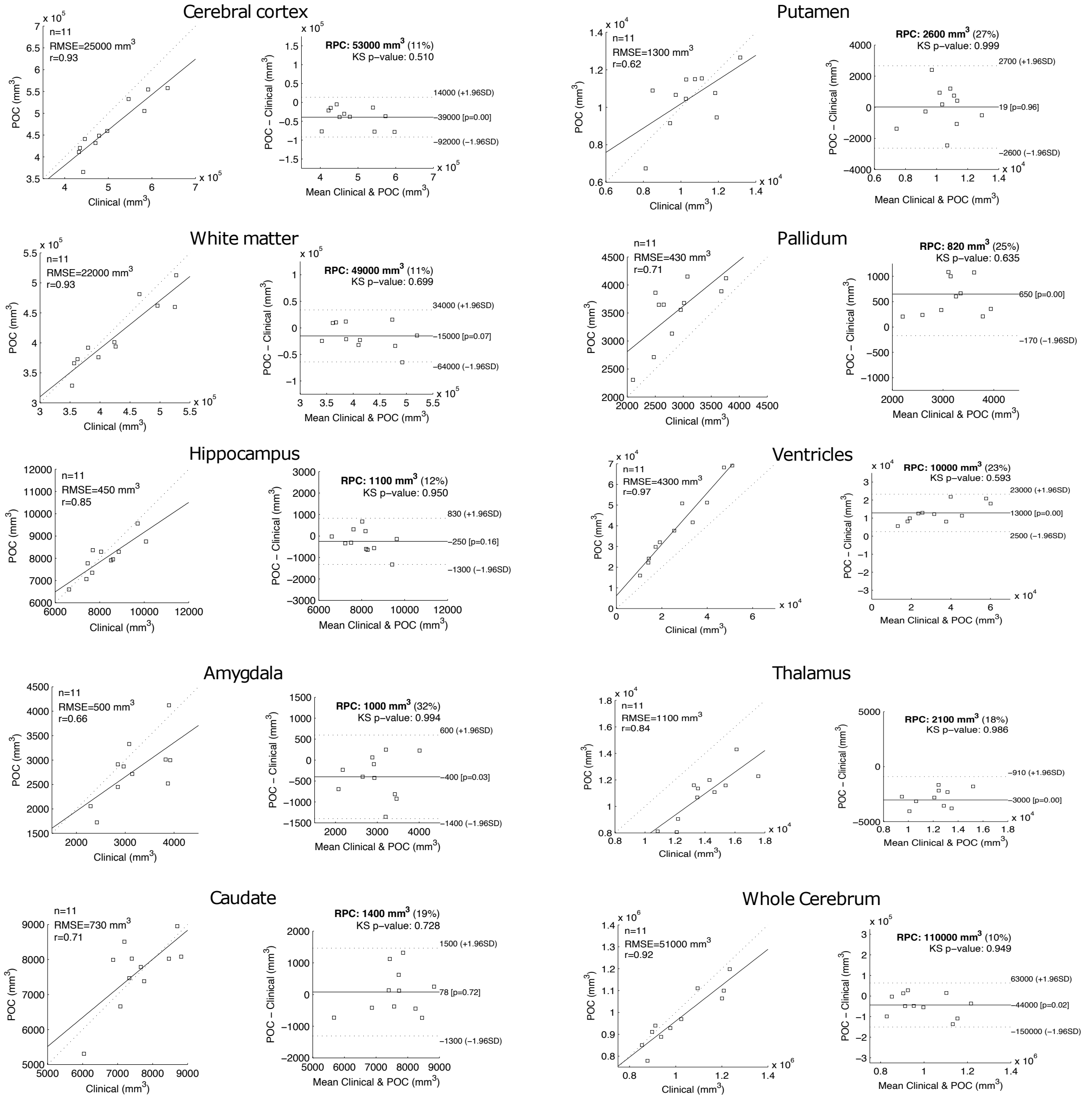

**Supplementary Figure 3 Scatter and Bland-Altman plots for our proposed method.** Scatter (left) and Bland-Altman plots (right) comparing the ROI volumes derived from the HF-MRI clinical scans (ground truth) and from the LF-MRI scans using our proposed method. In the Bland-Altmann plots, RPC stands for reproducibility coefficient, and the KS p-value is for a Kolmogorov-Smirnov test of normality of the differences. Please see caption of Supplementary Figure 1 for further details.